\documentclass{article}
\usepackage[T1]{fontenc}
\usepackage[utf8]{inputenc}
\usepackage{ismir} 
\usepackage{amsmath,cite,url}
\usepackage{graphicx}
\usepackage{color}
\usepackage{booktabs}
\usepackage{fontawesome5} 
\title{Equivariant Music Transformer}

\multauthor
  {Zixun Guo\textsuperscript{\faMusic} \hspace{1cm} Simon Dixon\textsuperscript{\faMusic}}
  {\textsuperscript{\faMusic}Centre for Digital Music, Queen Mary University of London, United Kingdom\\
  {\tt\small \{zixun.guo, s.e.dixon\}@qmul.ac.uk}
  }

\def\authorname{Z. Guo and S. Dixon}

\begin{document}

\maketitle

\begin{abstract}\label{sec:abstract}
Humans recognize a musical passage even when it is shifted in time or transposed in pitch, indicating a notion of equivariance in the representation space. Our analysis, however, shows that standard music transformers map such time-shifted or pitch-transposed inputs onto uncorrelated representations: these models become progressively less equivariant as they scale in size or train longer.
This suggests that in standard music transformers, additional model capacity is allocated to memorizing absolute patterns rather than capturing shared musical structures.
In this paper, we propose the Equivariant Music Transformer (EMT), which enforces equivariance through self-distillation by jointly optimizing a next-token-prediction and an auxiliary equivariance regularization loss. 
We find that the additional equivariance loss acts as a beneficial regularizer, simultaneously improving next-token prediction and producing equivariant latent representations.
Through both objective and subjective evaluations, EMT demonstrates superior equivariance and generative capability compared to data augmentation, feature engineering, and state-of-the-art (SOTA) baselines. 
More broadly, our findings reveal that standard language modeling methods alone do not capture music's translational symmetries, and dedicated inductive biases are required to produce better music representations. The code, weights and demos are available online\footnote{guozixunnicolas.github.io/equivariant-music-transformer-demo/}.

\end{abstract}

\section{Introduction}\label{sec:introduction}
Humans effortlessly recognize a musical passage regardless of whether it is transposed to a different key or shifted in time. This is because the perceived intervals remain unchanged even as the absolute pitches or onsets change. In computational terms, this corresponds to the notion of equivariance \cite{Cohen2016GroupEquivariant,Dowling1978ScaleContour}: a transformation of the input produces a predictable, corresponding transformation of the output. Incorporating the equivariant inductive bias has proven effective across a range of MIR tasks \cite{PESTO, Quinton22EquivariantTempoEstimation, Gagner24EquivariantTempoEstimation, Kong24Stone}. 

In the area of symbolic music, existing approaches either incorporate this inductive bias via input-level feature engineering, for instance, encoding symbolic music using relative attributes \cite{Guo23FME, Inaba24Relativity,guo2025moonbeam,Agarwal25FStripe,huang18musictransformer}, latent space modification \cite{huang18musictransformer,Guo23FME,guo2025moonbeam,melechovsky2024mustango} or data augmentation \cite{Donahue19LakhNES, Bradshaw25ScalingSymbolic}. Yet, none of these approaches enforces the equivariant property at the model level.

In this paper, we propose the Equivariant Music Transformer (EMT), which directly enforces equivariance in a generative transformer architecture via an auxiliary equivariance regularization loss, jointly optimized with  
  the standard next-token prediction objective.  We hypothesize that modifying the model's loss landscape is a stronger regularization signal than input-level feature engineering, and more generalizable than data augmentation. 

To our best knowledge, this is the first work that incorporates an equivariance regularization loss to the generative transformer architecture for music. We hence aim to answer the following research questions: 
1. Does the auxiliary equivariance loss act as a beneficial regularizer for next-token prediction, or does it introduce competing gradients that degrade generative performance?
2. How does the equivariance for both EMT and standard transformers evolve throughout training? And how does model-level regularization compare against data augmentation and feature-level equivariance regularization?
3. Finally, does equivariance emerge as transformer-based symbolic music models grow larger or train longer?

To answer the above questions, we measured the equivariance of the Anticipatory Music Transformer \cite{Thickstun24anticipatory_music_transformer}, a SOTA symbolic music model, and our own ablation model trained with the next-token-prediction loss only. We discover a notable trend: as models grow larger and train longer, they become progressively less equivariant. This suggests that standard transformers increasingly allocate representational and generative capacity to remember absolute patterns, mapping musically-equivalent inputs, such as onset-shifted or pitch-transposed versions of the same musical passage, onto distant, uncorrelated representations in the latent space. Our proposed EMT directly addresses this by constraining the latent space to map musically-shifted inputs onto correspondingly shifted representations, ensuring that the model's internal representations reflect the underlying musical structure.                              

From our objective evaluations, we find that incorporating equivariance regularization not only enforces the latent representation to be translation-equivariant, it also directly improves the transformer's generative capability, indicated by lower next-token-prediction losses. We attribute this improvement to a more efficient use of model capacity: rather than learning separate representations for each shifted variant, the model has learnt to link them as correlated representations. Moreover, under identical transform distributions and
  compute budgets, our proposed model-level regularization also outperforms data augmentation in both generative capability and generalization to unseen musical shifts. Additionally, we find that our proposed method is complementary to feature-level equivariance, implemented via fundamental music embedding (FME) \cite{Guo23FME} and multi-dimensional relative attention (MRA) \cite{guo2025moonbeam}\footnote{FME is a music-distance-aware input embedding layer and MRA replaces standard RoPE: it encodes relative musical positions (e.g., relative onsets and pitches), instead of relative token positions.}, with their combination yielding the best results.
  We also conduct a listening test to evaluate the robustness of symbolic music transformers under musically-shifted input prompts. Our results indicate that the baseline model exhibits a larger degradation in both smoothness and     
  enjoyment scores under shifted conditions, while our model maintains more consistent quality. Our contributions can be summarized as follows: 
\begin{enumerate} 
  \item We reveal that standard symbolic music transformers become progressively less equivariant as they scale or train for more iterations. To the best of our knowledge, this provides the first empirical evidence that equivariance does not emerge from scaling. 
  \item We propose the Equivariant Music Transformer (EMT), a transformer architecture jointly optimized with next-token prediction and equivariance regularization objectives. The auxiliary loss not only produces music representations that align with human perception of music, but also directly improves the model's generative capabilities.
  \item EMT achieves substantially stronger equivariance and better generative capabilities compared to data augmentation, feature-level equivariance, and SOTA external baselines. 
  \item Through a listening test, EMT achieves comparable unconditional music generation quality compared to the Anticipatory Music Transformer \cite{Thickstun24anticipatory_music_transformer}, despite being a smaller model. Moreover, we confirm that EMT maintains consistent generative quality and smoothness under musically-shifted prompts, while the baseline exhibits more significant degradations.                                                 
\end{enumerate}  

\section{Related Work}

A model is \textit{equivariant} to a transformation if the transformation at the input results in a corresponding, predictable transformation at the output. For example, transposing a melody by 2 semitones should shift the output pitch distribution by the same amount. Formally, let $\mathcal{T}$ denote a transformation in the input space and $\mathcal{T}'$ be its corresponding transformation in the output space. A model $f$ is equivariant if $f(\mathcal{T}(x)) = \mathcal{T'}(f(x))$. In the context of music, equivariance provides an important inductive bias, as many MIR tasks rely on the model’s ability to model underlying tonal or translational symmetries of music.

Consequently, equivariance can be treated as a supervision signal for music representation learning, and has proven to be effective in many tasks including: $f_0$, tempo, and tonality estimation \cite{PESTO, Quinton22EquivariantTempoEstimation, Gagner24EquivariantTempoEstimation, Kong24Stone}, music representation learning \cite{Guinot25Leov}, as well as synthesizer inversion \cite{Hayes25synthesizerinversion}. Furthermore, equivariance metrics are integrated into evaluation benchmarks \cite{Plachouras25Benchmark} to indicate the model's ability to capture the underlying data symmetries. Most of these prior works apply an equivariance supervision loss on static feature encoders, and its impact on the autoregressive transformer architecture remains unexplored.

Hypothetically, including an equivariance regularization in the autoregressive transformer architecture ensures that the generative model does not waste representational capacity on modeling different views of the same data. However, a key open question is whether the regularization will produce conflicting gradients, hence affecting the model's generative capability. Existing work encourages approximate-equivariance via input-level feature engineering \cite{Guo23FME, luo24music102, guo2025moonbeam} (e.g., replace standard embedding layers with musical-distance-aware embeddings), architectural change \cite{luo24music102, Lattner18TranspositionInvariant, Lattner18Predictive, Guo23FME, guo2025moonbeam, huang18musictransformer, Inaba24Relativity}, latent-space modification \cite{Guo23FME, Agarwal25FStripe, Inaba24Relativity, huang18musictransformer} (e.g., include relative position embeddings), or data augmentation \cite{Donahue19LakhNES, Bradshaw25ScalingSymbolic}. Generally speaking, these methods encourage the autoregressive model to recognize patterns regardless of their absolute token positions.

While these approaches consistently outperform their unconstrained counterparts, their training objectives almost always default to a standard next-token cross-entropy loss. We hypothesize that, despite the effectiveness of feature-level engineering, incorporating explicit loss- and model-level equivariance provides a stronger supervision signal, hence enabling the model itself to become equivariant.

\section{Method}\label{sec:method}
\begin{figure*}[htbp]
\centering
\includegraphics[width=0.7\textwidth]{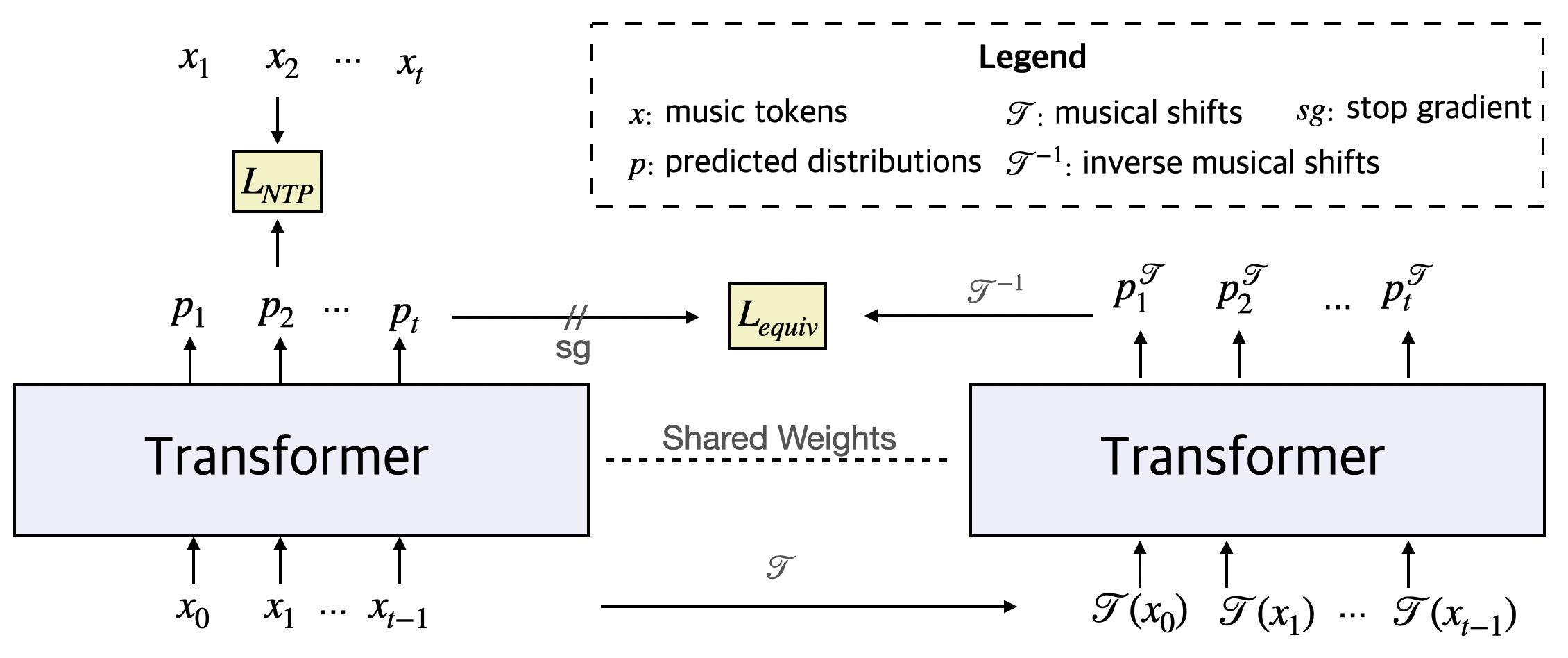}
\caption{The main branch (left) performs next-token prediction; the auxiliary branch (right) self-distills the main branch, aligning its prediction on shifted inputs with the correspondingly-shifted original distribution to enforce 
equivariance.}
\label{fig:architecture}
\end{figure*}
Figure \ref{fig:architecture} outlines the architecture of our proposed Equivariant Music Transformer. The main branch to the left represents a generative transformer, which is trained to autoregressively predict the music token $x_{t}$ at the next step $t$. More specifically, given a music sequence $x_0$ … $x_t$ with length $t+1$, the transformer uses a cross-entropy loss as the optimization target during training:
\begin{equation}\label{eq:ntp}
\mathcal{L}_\mathit{NTP} = - \frac{1}{t} \sum_{i=1}^{t} \log p(x_i \mid x_0, \ldots, x_{i-1})
\end{equation}
where $p$ represents the predicted probability distribution.

The auxiliary branch to the right in Figure \ref{fig:architecture} distills from the main branch to enforce equivariant representations. To calculate the auxiliary loss, we apply a random musical shift $\mathcal{T}$ to the music sequence $x$. This transformation $\mathcal{T}$ comprises random pitch and temporal shifts, which are applied additively to the input $x$. The shifted music sequence $\mathcal{T}(x)$ is then input to the shared-weight transformer, yielding output probability distributions $p_i^{\mathcal{T}}$ for each step $i$. 

Supposing that the model is equivariant, the probability distribution $p^{\mathcal{T}}$ should be equivalent to $p$ but shifted according to $\mathcal{T}$. In other words, by applying the inverse transform $\mathcal{T}^{-1}(p^{\mathcal{T}})$, we could obtain a distribution identical to $p$, if the model is equivariant. Here $\mathcal{T}^{-1}$ is implemented in torch via torch.roll with necessary masking at logit boundaries. Hence, the auxiliary equivariance loss is implemented using KL divergence as follows:
\begin{equation}\label{eq:equiv}
\mathcal{L}_\mathit{equiv} = \frac{1}{t} \sum_{i=1}^{t} \text{KL} \left( \text{sg}(p_i) \parallel \mathcal{T}^{-1}(p_i^{\mathcal{T}}) \right)
\end{equation}
Note that computing $\mathcal{L}_\mathit{equiv}$ requires a second forward pass through the shared-weight transformer, approximately doubling the per-step training cost. Here, we apply a stop-gradient operation, denoted as $\text{sg}(\cdot)$ in Equation \ref{eq:equiv}, to the anchor predictions $p_i$. This ensures that the main branch acts as a stable next-token prediction model, forcing the shifted branch to map its outputs to the model's best current understanding of the unshifted music. Without this stop-gradient operation, the model could potentially bypass learning true equivariance by collapsing both branches into a trivial distribution.

Finally, the model jointly optimizes a combined objective function consisting of the standard autoregressive next-token prediction ($\mathcal{L}_\mathit{NTP}$) loss and the auxiliary equivariance loss ($\mathcal{L}_\mathit{equiv}$) during training.
\begin{equation}\label{eq:total}
\mathcal{L}_\mathit{total} = \mathcal{L}_\mathit{NTP} + \lambda \mathcal{L}_\mathit{equiv}
\end{equation}
Here $\lambda$ is the hyperparameter controlling the strength of the equivariance regularization. 

\subsection{Transformer Architecture}
\begin{figure}[htbp]
\centering
\includegraphics[width=0.3\textwidth]{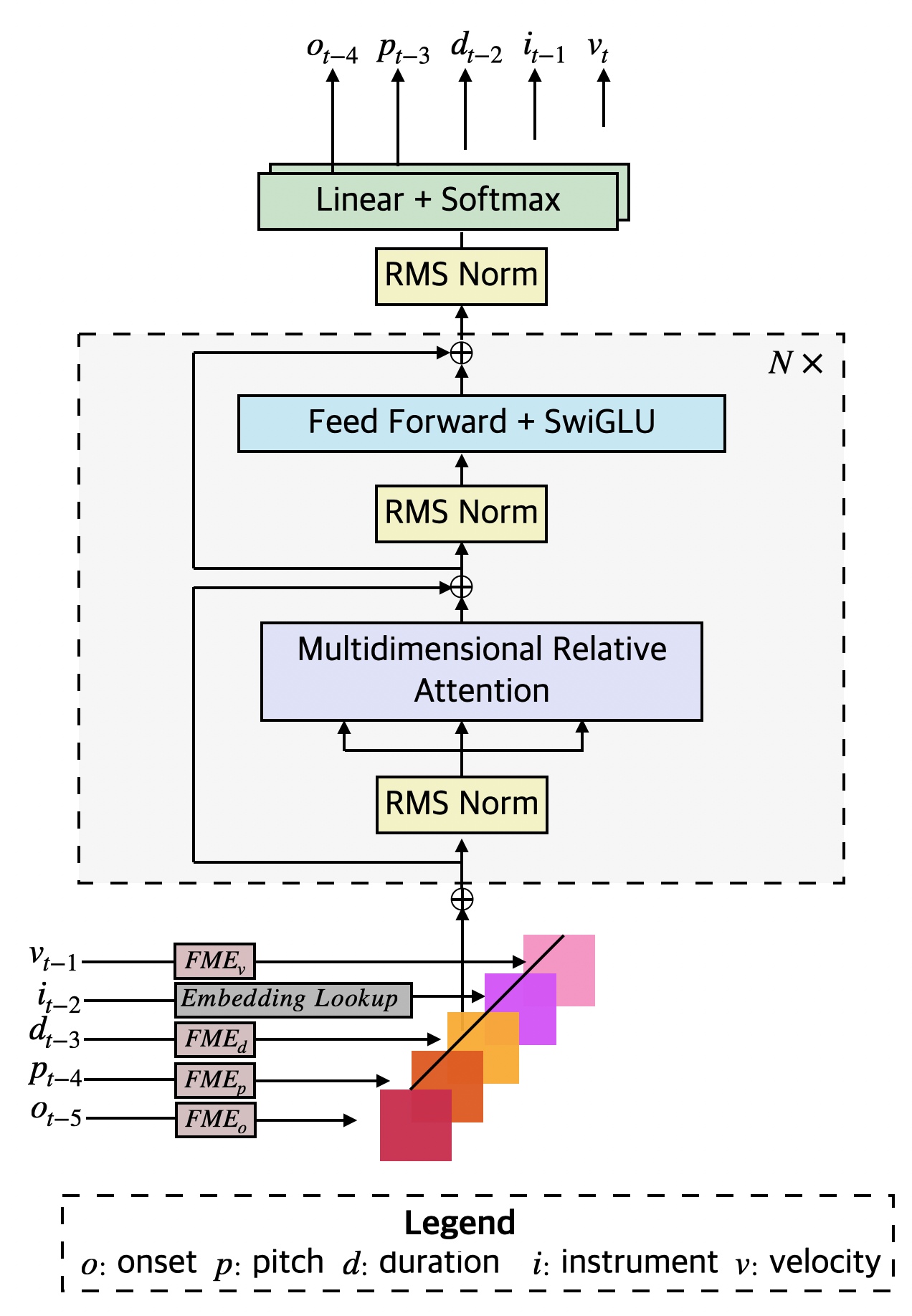}
\caption{Modified Moonbeam \cite{guo2025moonbeam} architecture (a generative transformer) with delayed inputs and outputs.}
\label{fig:transformer_architecture}
\end{figure}

We base EMT on the transformer used in Moonbeam \cite{guo2025moonbeam}, with several key modifications (see Figure \ref{fig:transformer_architecture}) to facilitate our dual-objective loss formulation. We chose this particular architecture because it implicitly incorporates feature-level equivariance through its Fundamental Music Embedding (FME) \cite{Guo23FME}. Furthermore, it also encodes relative musical positions (e.g., pitch and onset intervals), instead of relative token positions, in the proposed multi-dimensional relative attention \cite{guo2025moonbeam}. Using this foundation allows us to isolate the effects of model-level (loss-driven) versus feature-level equivariance easily.

Next we describe the modifications we applied to the original architecture, shown in Figure \ref{fig:transformer_architecture}: First, to facilitate the formulation of $\mathcal{L}_\mathit{equiv}$, we predict absolute rather than relative onsets, and we merge the previously separated octave and pitch-class tokens into a single pitch dictionary. These adjustments ensure that our sampled musical transformations ($\mathcal{T}$) can be applied across the input and output space easily. Second, we simplify the original architecture by replacing the original GRU decoder with a token delay pattern to capture the local inter-dependencies between different musical attributes, similar to \cite{Copet23musicgen, wang25timeshifedtoken}. This technique effectively infuses attribute correlation without adding a separate decoder. Furthermore, we simply add rather than concatenate the input embeddings. This allows each attribute to have larger embedding dimensions, increasing its expressiveness. 

\section{Experiment Settings}\label{sec:experiments}

\subsection{Dataset and Preprocessing}
Our models are trained using different subsets of the LakhMIDI dataset \cite{raffel2016lakh}. We conduct our ablation studies using the clean subset (LMD clean) due to computation constraints. We randomly hold out 10\% of the clean subset as the test data, and use the rest for training. For subjective evaluations, we train our model on the entire LakhMIDI dataset (LMD-full) to compare against the baseline model: Anticipatory Music Transformer \cite{Thickstun24anticipatory_music_transformer}, following the same train-test split. Note that our model is $\sim$1.3 times smaller than the smallest baseline model.

To ensure high data quality, we filter the dataset using the following criteria: we remove files that are either excessively long or short (longer than 1 hour or less than 5 seconds), or contain inter-onset intervals (IOI) exceeding 60 seconds. The filtered MIDI files are then converted into a sequence of music events consisting of 5-way event tuples: $x = (o, p, d, i, v)$ (Figure \ref{fig:transformer_architecture}), which represent onset, pitch, duration, instrument, and velocity, respectively. Special \texttt{<SOS>} and \texttt{<EOS>} (Start, End of Sequence) tokens are prepended and appended to each sequence. 

To format the data for autoregressive modeling, we apply the delay pattern mentioned in Section \ref{sec:method} and continuously concatenate the sequences into fixed-length blocks of $L_\mathit{seq} = 1024$, a standard technique in language modeling \cite{Llama3}. Within these blocks, onset values are relativized for each segmented chunk, following the implementation of \cite{Thickstun24anticipatory_music_transformer}. Moreover, an attention mask is constructed for each sequence to avoid attention leakage across different concatenated tracks. For the clean subset, this yields approximately 79k training sequences and 8.7k testing sequences, corresponding to roughly 80M and 8.9M tokens for training and testing, respectively. Finally, to optimize I/O throughput during training, all preprocessed data are serialized and saved as memory-mapped files. 

\subsection{Training Setup} \label{sec:training_setup}
To increase training efficiency, the model training is distributed across 2 NVIDIA A100 or H100 GPUs, utilizing PyTorch's Distributed Data Parallel (DDP) framework with the NCCL backend.  Moreover, to maximize the utilization of NVIDIA Tensor Cores, we employed mixed precision training with the \texttt{bfloat16} data type, selectively upcasting to \texttt{fp32} for numerically sensitive operations to avoid precision loss (e.g., FME \cite{Guo23FME}, RoPE, LayerNorm, SoftMax). 

All models have 12 layers, with a hidden size of 768 and an intermediate size of 2048. This accounts for 96M total trainable parameters. For models utilizing equivariance regularization, the loss weight $\lambda$ is set to $0.001$. The value is selected empirically since the equivariance loss (KL divergence) is numerically larger than the NTP loss, requiring a smaller weight to avoid dominating the optimization. We use the AdamW optimizer with $\beta_1 = 0.9$, $\beta_2 = 0.95$, and a weight decay of $0.1$ and a learning rate scheduler with a 3\% linear warmup to a peak learning rate of $3 \times 10^{-4}$, followed by a cosine annealing decay down to a minimum learning rate set at 10\% of the peak. 

For models incorporating auxiliary losses, we dynamically sample musical transformations during training. Pitch shifts are constrained to a range of $[-12, 12]$ semitones and we allow time shifts ranging from $0-5$ seconds in $0.25$-second steps. From the resulting pool of 44 single (24 pitch shifts and 20 time shifts) and 480 combined transforms (applying pitch and time shifts simultaneously), we uniformly sample a single shift 50\% of the time and a paired combination the remaining 50\% of the time.  

\subsection{Ablation Studies and Baseline Models}
To study the behaviour of feature- and model-level equivariance regularization, we define several internal models and external baselines: (1) EMT without feature-level equivariance (standard embedding layer and RoPE), optimized with $\mathcal{L}_\mathit{NTP}$ only. (2) Model 1 with feature-level equivariance (See Section \ref{sec:introduction} for details) (3) Model 2 with an auxiliary data augmentation branch computing $\mathcal{L}_\mathit{NTP}$ instead of $\mathcal{L}_\mathit{equiv}$ on shifted inputs (architecturally identical to Figure \ref{fig:architecture}) (4) Model 1 but optimized with $\mathcal{L}_\mathit{NTP}$ and the auxiliary $\mathcal{L}_\mathit{equiv}$ (5) Our proposed EMT trained with both $\mathcal{L}_\mathit{NTP}$ and the auxiliary $\mathcal{L}_\mathit{equiv}$, with feature-level equivariance. 
For all models requiring an auxiliary loss (Model 3-5), we utilized a fixed random seed to ensure they are exposed to identical transformations during training. We also benchmark equivariance level (see Section \ref{sec:evaluation_setup}) against 2 external models: Anticipatory Music Transformer \cite{Thickstun24anticipatory_music_transformer} and MIDI-LLM \cite{wu2025midillm}.

\subsection{Objective and Subjective Evaluation Settings} \label{sec:evaluation_setup}
For objective evaluations, we evaluate the models across 2 dimensions: generative capability measured using validation cross-entropy loss across all attributes, and equivariance level, measured as follows. We first sample from a pool of musical transforms that is a superset of the musical transforms used during training (see Section \ref{sec:training_setup}) to thoroughly test the models' generalization to out-of-distribution shifts. We limit the pitch shifts within $[-24, 24]$ semitones and onset shifts from $0$ to $10$ seconds at $0.1$-second intervals. From this extended pool, we randomly sampled 20\% of all single shifts (48 pitch and 100 time shifts) and 1\% of all combined shifts (4800 combinations), totaling 79 evaluation musical shifts.  We then report the equivariance level using: $\mathcal{L}_\mathit{equiv}$ (See Section \ref{sec:method}), Top-1 matching accuracy, and Top-5 Jaccard similarity for both shifted and unshifted attributes. Top-1 matching accuracy measures the match rate of the highest-probability token between the original and inversely shifted output distributions. Top-5 Jaccard similarity computes the Intersection over Union between the unranked sets of the 5 most likely tokens from both distributions.


Finally, to subjectively assess generative quality and robustness to input transformations, we conduct a Mean Opinion Score (MOS) listening test. For both our proposed model and the baseline,  we prompt the models using two sets of ten 5-second prompts, limiting output duration to 30 seconds. The second set of prompts consists of time- or pitch-shifted versions of the first. These transformations consist of a pitch transposition of up to ±6 semitones and a minor temporal onset shift up to 1 second, which sometimes can be nearly unnoticeable to the participants. This yields a total of twenty 30-second musical excerpts. We recruited 14 participants, of whom 12 passed a screening task (>75\% accuracy on identifying smooth vs.\ unsmooth transitions) and are included in the final analysis.   

During the test, participants are provided with the audio alongside a synchronized piano roll visualizer that explicitly highlights the prompt region, allowing them to clearly contextualize the model's continuation. To eliminate presentation bias, the order of the generated tracks is randomized for each participant and they are asked to rate two specific criteria: (1) the smoothness of the transition between the prompt and the generated sequence, and (2) their overall musical preference. In our results, we report the MOS for both the shifted and unshifted conditions across both models. We also report the change in MOS ($\Delta$MOS) between the unshifted and shifted conditions, serving as a metric of each model's robustness to input transformations.

\section{Results}\label{sec:results}

\begin{table*}[t]
\centering
\resizebox{\textwidth}{!}{
\begin{tabular}{l cc cccccc ccc ccc ccc}
\toprule
& & & \multicolumn{6}{c}{\textbf{$\mathcal{L}_\mathit{NTP}$ $\downarrow$}} & \multicolumn{3}{c}{\textbf{$\mathcal{L}_\mathit{equiv}$ $\downarrow$}} & \multicolumn{3}{c}{\textbf{Top-1 Accuracy $\uparrow$}} & \multicolumn{3}{c}{\textbf{Top-5 Jaccard $\uparrow$}} \\
\cmidrule(lr){4-9} \cmidrule(lr){10-12} \cmidrule(lr){13-15} \cmidrule(lr){16-18}
\textbf{Model} & \textbf{Reg.} & \textbf{Feat. Equiv.} & All & Onset & Pitch & Dur. & Inst. & Vel. & All & Shift & Unshift & All & Shift & Unshift & All & Shift & Unshift \\
\midrule
\multicolumn{18}{l}{\textit{Internal Models}} \\
\midrule
\textbf{1} & $\times$ & $\times$ & 0.912 & 0.747 & 0.913 & 1.334 & 0.140 & 1.426 & 0.286 & 0.502 & 0.182 & 0.820 & 0.750 & 0.853 & 0.566 & 0.492 & 0.594 \\
\textbf{2} & $\times$ & FME+MRA & 0.889 & 0.704 & 0.890 & 1.316 & 0.137 & 1.396 & 0.290 & 0.558 & 0.167 & 0.709 & 0.604 & 0.755 & 0.605 & 0.521 & 0.636 \\
\textbf{3} & Data Aug. & FME+MRA & 0.853 & 0.650 & 0.838 & 1.277 & \textbf{0.126} & 1.374 & 0.069 & 0.134 & 0.039 & 0.901 & 0.879 & 0.910 & 0.704 & 0.646 & 0.724 \\
\textbf{4} & Equiv. Loss & $\times$ & 0.849 & 0.654 & 0.830 & 1.283 & 0.130 & \textbf{1.350} & 0.077 & 0.152 & 0.040 & 0.901 & 0.873 & 0.914 & 0.694 & 0.642 & 0.713 \\
\textbf{5}(Our proposed EMT) & Equiv. Loss & FME+MRA & \textbf{0.847} & \textbf{0.636} & \textbf{0.823} & \textbf{1.275} & 0.127 & 1.373 & \textbf{0.049} & \textbf{0.104} & \textbf{0.022} & \textbf{0.917} & \textbf{0.891} & \textbf{0.928} & \textbf{0.747} & \textbf{0.677} & \textbf{0.775} \\
\midrule
\multicolumn{18}{l}{\textit{External Baselines}} \\
\midrule
Anticipatory (S, 100K) & $\times$ & $\times$ & $-$ & $-$ & $-$ & $-$ & $-$ & $-$ & 0.238 & 0.364 & 0.094 & 0.871 & 0.864 & 0.869 & 0.650 & 0.541 & 0.766 \\
Anticipatory (S, 800K) & $\times$ & $\times$ & $-$ & $-$ & $-$ & $-$ & $-$ & $-$ & 0.292 & 0.393 & 0.186 & 0.855 & 0.863 & 0.832 & 0.620 & 0.538 & 0.704 \\
Anticipatory (M, 100K) & $\times$ & $\times$ & $-$ & $-$ & $-$ & $-$ & $-$ & $-$ & 0.397 & 0.565 & 0.222 & 0.835 & 0.821 & 0.836 & 0.594 & 0.483 & 0.703 \\
Anticipatory (M, 200K) & $\times$ & $\times$ & $-$ & $-$ & $-$ & $-$ & $-$ & $-$ & 0.654 & 0.973 & 0.311 & 0.787 & 0.762 & 0.795 & 0.537 & 0.422 & 0.649 \\
Anticipatory (M, 800K) & $\times$ & $\times$ & $-$ & $-$ & $-$ & $-$ & $-$ & $-$ & 0.741 & 1.043 & 0.406 & 0.765 & 0.730 & 0.785 & 0.522 & 0.409 & 0.629 \\
Anticipatory (L, 100K) & $\times$ & $\times$ & $-$ & $-$ & $-$ & $-$ & $-$ & $-$ & 0.495 & 0.658 & 0.338 & 0.826 & 0.819 & 0.813 & 0.567 & 0.466 & 0.656 \\
Anticipatory (L, 800K) & $\times$ & $\times$ & $-$ & $-$ & $-$ & $-$ & $-$ & $-$ & 0.813 & 1.159 & 0.444 & 0.790 & 0.759 & 0.806 & 0.526 & 0.416 & 0.619 \\
MIDI-LLM \cite{wu2025midillm} & $\times$ & $\times$ & $-$ & $-$ & $-$ & $-$ & $-$ & $-$ & 0.394 & 0.537 & 0.237 & 0.812 & 0.812 & 0.798 & 0.539 & 0.472 & 0.602 \\
\bottomrule
\end{tabular}
}
\caption{Evaluation results. The next-token prediction loss ($\mathcal{L}_\mathit{NTP}$) reflects generative capability; the equivariance loss ($\mathcal{L}_\mathit{equiv}$), Top-1 Accuracy, and Top-5 Jaccard metrics reflect structural robustness. \textit{Shift} and \textit{Unshift} denote metrics calculated on the shifted and unshifted attributes. Feature-level equivariance is implemented via FME \cite{Guo23FME} and MRA \cite{guo2025moonbeam}.}
\label{tab:comprehensive_results}
\end{table*}

\subsection{Equivariance and Loss Evolution}

\begin{figure}[tbp]
\centering
\includegraphics[width=0.5\textwidth]{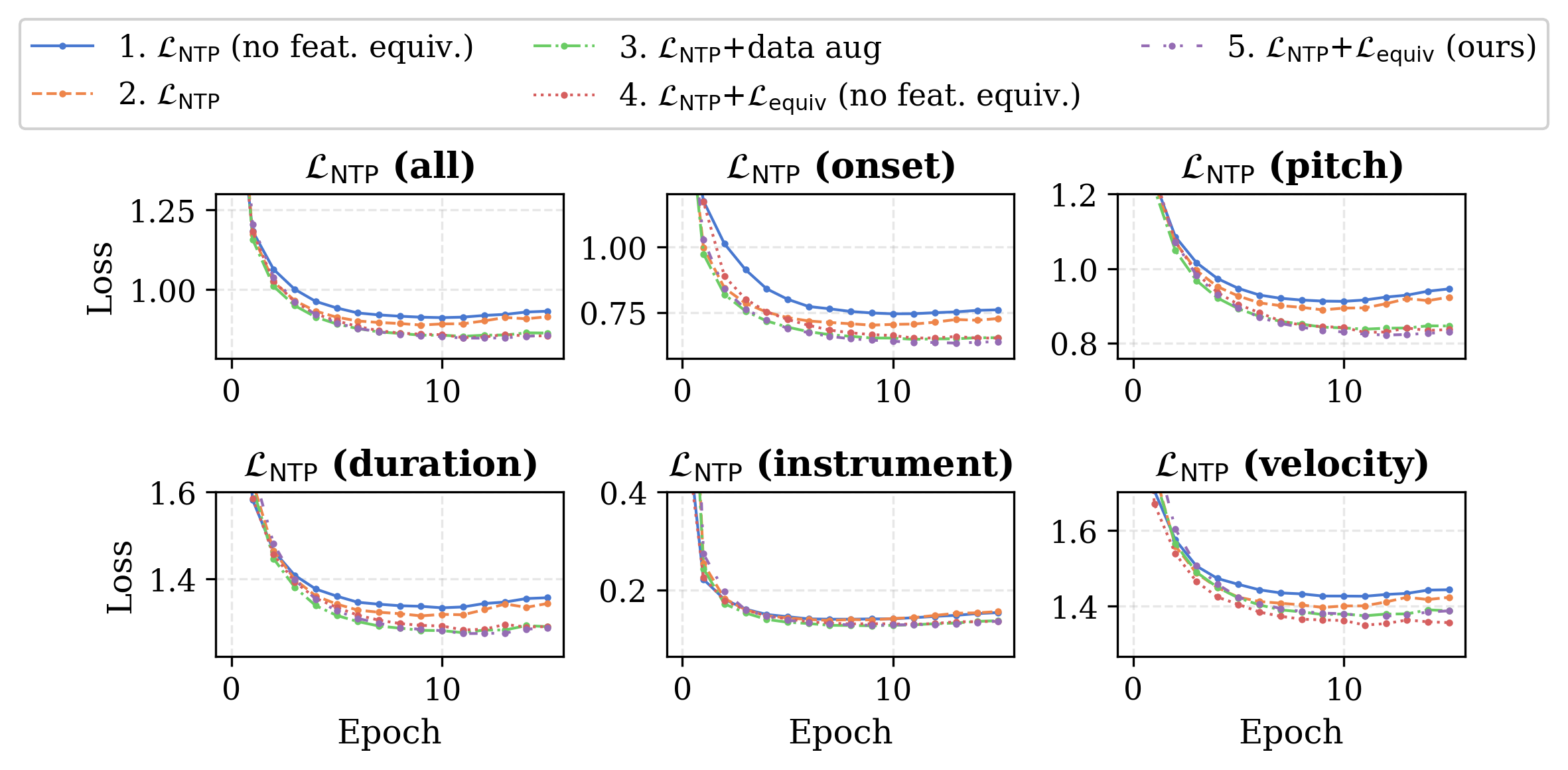}
\caption{Validation $\mathcal{L}_\mathit{NTP}$ throughout training for all internal model variants, with EMT achieving the lowest loss.}
\label{fig:loss_evolution}
\end{figure}

\begin{figure}[tbp]
\centering
\includegraphics[width=0.5\textwidth]{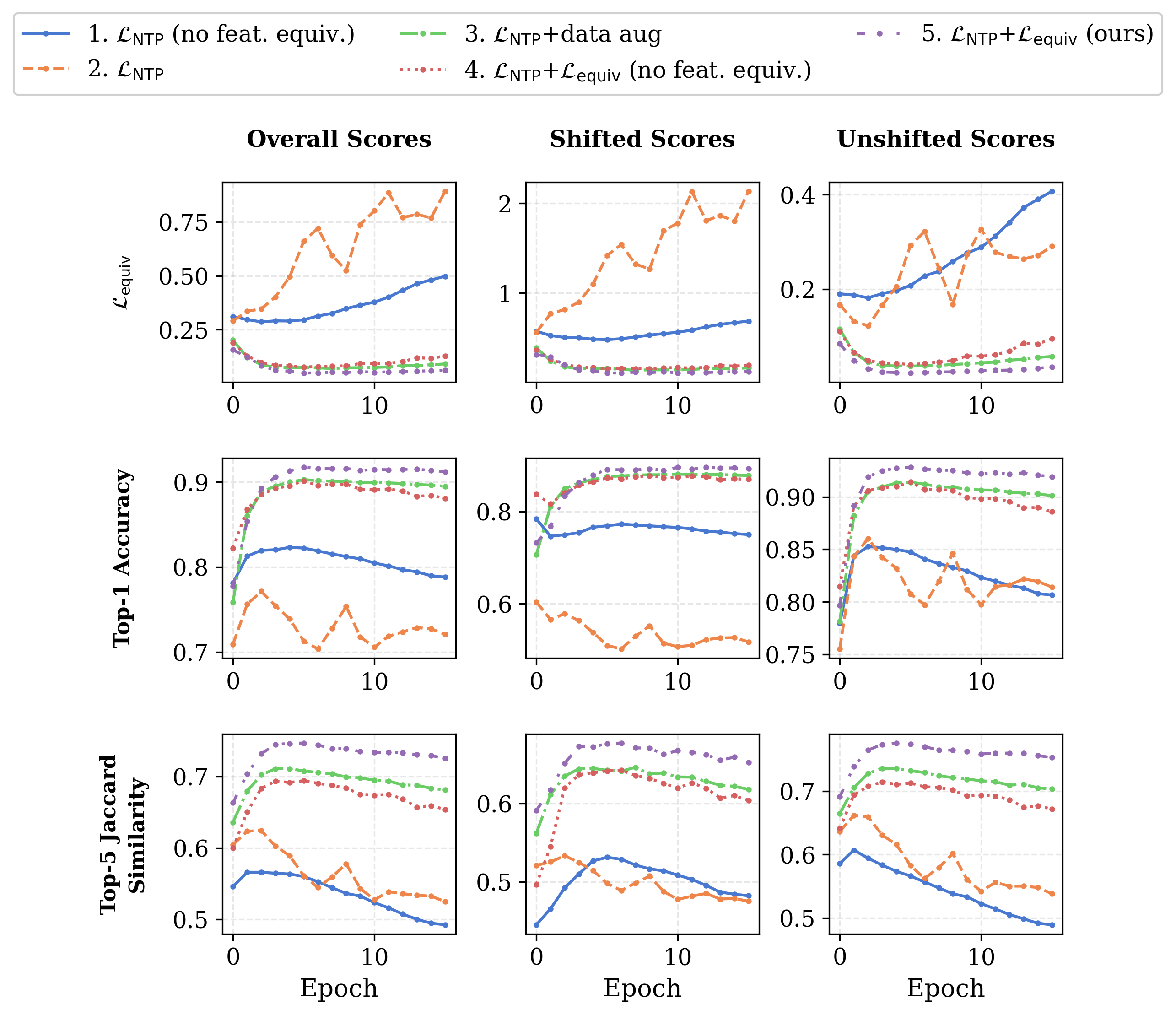}
\caption{Equivariance metrics throughout training.  Models (1 and 2) trained only with $\mathcal{L}_\mathit{NTP}$ show degrading equivariance, opposite to the regularized models.}
\label{fig:equiv_evolution} 
\end{figure}

\textbf{Next-token Prediction Loss:} Throughout training, the validation loss across all models consistently decreases before consolidating into stable convergence (Figure \ref{fig:loss_evolution}). Overall, models incorporating auxiliary signals (Model 3-5) achieve lower validation losses than the models trained only with $\mathcal{L}_\mathit{NTP}$ (Models 1-2). This shows that explicitly adding an equivariance regularizer not only enforces structural coherence but also directly improves the model's generative capabilities. Among the regularized models, our proposed EMT, which incorporates both feature- and model-level equivariance regularization, achieves the lowest loss, followed by Model 3 (data augmentation) and Model 4 (no feature equivariance). While these results confirm that both data augmentation and our proposed equivariance loss function operate as effective regularizers, the combination of model- and feature-level constraints proved to be the best approach. Furthermore, the loss curve remained highly stable, suggesting potential for future scalability.

\textbf{Equivariance Level:} In contrast to the uniform trend of the validation loss throughout training, the equivariance metrics show diverging behaviours among different types of models (Figure \ref{fig:equiv_evolution}).  For Model 1-2 trained with only $\mathcal{L}_\mathit{NTP}$, the equivariance metrics exhibited continuous degradation as training progressed, despite their continuously improving validation losses. This indicates that these models try to memorize temporally or pitch-shifted music sequences as entirely distinct inputs rather than recognizing their common underlying patterns, hence failing to generalize to out-of-distribution data.

On the contrary, Model 3-5 show the opposite behavior, with improving equivariance metrics throughout training. Among these, EMT demonstrates the strongest equivariance properties. Model 3, which uses data augmentation trained with identical musical transforms, yielded noticeably weaker generalizability, indicated by higher $\mathcal{L}_\mathit{equiv}$ and lower Top-1 matching accuracy and Top-5 Jaccard similarity. This indicates that data-level regularization (augmentation) generalizes worse than our proposed model-level equivariance regularization. We hypothesize that $\mathcal{L}_\mathit{equiv}$ enables the model to recognize how much the input is shifted, while the augmentation-only model treats shifted inputs as independent samples without modelling the shift itself. Finally, Model 4, which is optimized using model-level equivariance loss but without careful feature engineering, yielded the worst equivariance among the regularized models. This is unsurprising since vanilla RoPE only informs the model about relative token positions but does not provide useful relative music information, as opposed to \cite{guo2025moonbeam}. This indicates that even with model-level regularization, the model still benefits from domain inductive biases. To summarize, given identical computation budgets and transforms, combining model- and feature-level equivariance regularization is the most effective. 

\subsection{Equivariance Comparison against Baselines}
In Table \ref{tab:comprehensive_results}, when comparing against external baselines, EMT achieved the best overall equivariance across all metrics. Notably, our method achieves significantly lower $\mathcal{L}_\mathit{equiv}$. This means musically-shifted inputs are mapped to correspondingly shifted representations in the latent space. On the contrary, since the baseline models are optimized by $\mathcal{L}_\mathit{NTP}$ only, and the input representation space (i.e., standard embedding layers, position-based RoPE) does not preserve the relative music information, musically-shifted inputs are treated as entirely different token sequences and representations. As a result, the representations of musically-shifted inputs are mapped onto uncorrelated representations in the latent space, as shown by higher $\mathcal{L}_\mathit{equiv}$, lower Top-1 Accuracy and Top-5 Jaccard Similarity.

Since the Anticipatory model \cite{Thickstun24anticipatory_music_transformer} provides checkpoints across 3 model sizes (S, M, L) at multiple training stages (100K and 800K steps), we can analyze how the representation space of a standard transformer evolves with scale and training duration. We expect this trend to hold broadly for symbolic music transformers optimized with $\mathcal{L}_\mathit{NTP}$. The results reveal a clear trend: 
  larger models and longer training both lead to decreased equivariance. For instance, the medium model's $\mathcal{L}_\mathit{equiv}$ increases from 0.397 at 100K steps to 0.654 and 0.741 at 200K and 800K steps. Similarly, at 800K steps, equivariance degrades with model size:           
  $\mathcal{L}_\mathit{equiv}$ rises from 0.292 (S) to 0.741 (M) to 0.813 (L). This trend is consistent with our internal ablation models (Models 1-2 in Figure \ref{fig:equiv_evolution}). This suggests that equivariance does not emerge from scaling. Instead, without explicit equivariance supervision, transformers increasingly allocate capacity to memorizing absolute patterns rather than recognizing the shared underlying structure.

\subsection{Listening Test Results}

\begin{table}[htbp]
\centering
\resizebox{0.8\columnwidth}{!}{
\begin{tabular}{l cc c cc c}
\toprule
& \multicolumn{3}{c}{\textbf{Smoothness}} & \multicolumn{3}{c}{\textbf{Enjoyment }} \\
\cmidrule(lr){2-4} \cmidrule(lr){5-7}
\textbf{Model} & Unshift $\uparrow$ & Shift $\uparrow$& $\Delta MOS$ $\downarrow$ & Unshift $\uparrow$ & Shift $\uparrow$ & $\Delta MOS$ $\downarrow$\\
\midrule
 EMT & \textbf{3.18 $\pm$ 1.40} & \textbf{2.98 $\pm$ 1.43} & \textbf{0.20} & \textbf{3.02 $\pm$ 1.24} & \textbf{2.72 $\pm$ 1.11} & \textbf{0.30} \\
  Anticipatory (M) & 2.68 $\pm$ 1.24 & 2.10 $\pm$ 1.09 & 0.58 & 2.98 $\pm$ 1.00 & 2.25 $\pm$ 0.93 & 0.73 \\
\midrule
  $p$ (Wilcoxon) & & & 0.039 & & & 0.008 \\
\bottomrule
\end{tabular}
}
\caption{Listening test results. $\Delta$ denotes the drop from unshifted to shifted conditions, indicating robustness.}
\label{tab:listening_test}
\end{table}

The listening test results in Table \ref{tab:listening_test} show that the $\Delta MOS$ between unshifted and shifted conditions is substantially smaller for EMT across both smoothness and enjoyment criteria. A paired Wilcoxon signed-rank test confirms that the difference in $\Delta MOS$ is statistically significant. This indicates that EMT maintains consistent generative quality under musically-shifted prompts, proving the effectiveness of explicit equivariance regularization. For the Anticipatory model, we observe that when it is prompted with musically-shifted inputs, the model initially generates incoherent continuations before gradually recovering as generation progresses further from the prompt. We attribute this to the model treating the shifted prompt as out-of-distribution data. We invite readers to listen to samples from both models here\footnote{guozixunnicolas.github.io/equivariant-music-transformer-demo}. This phenomenon is reflected by its large $\Delta MOS$ score. 
Moreover, despite being a smaller model, EMT achieves comparable unconditional generation quality compared to the Anticipatory baseline, reflected by the unshifted enjoyment score (3.02 vs 2.98). 

  \section{Conclusion}                                                                                                                                                                                                                               
                      
  We proposed the Equivariant Music Transformer (EMT), jointly optimized with the next-token prediction and equivariance regularization loss.
Through both subjective and objective evaluations, we show that adding the equivariance inductive bias not only produces music representations that align with human perception of music better, but also improves the transformer's music generation capability.
EMT also achieves better equivariance and generative capability compared to strong baselines, while maintaining consistent quality under musically-shifted                                                                                                                                                           
inputs. This highlights the importance of latent space regularization, enabling the model to use its capacity more efficiently.

\section{Acknowledgments}

Zixun Guo is a PhD student at the UKRI Centre for Doctoral Training (CDT) in Artificial Intelligence and Music (AIM), supported by UK Research and Innovation [grant number EP/S022694/1].
\section{AI Usage Statement}

We used generative AI assistants (Gemini and Claude) to improve the grammar, spelling, and phrasing of the authors' original text, and to help format \LaTeX{} tables. We also used Claude Code to assist in implementing the codebase, all of which was manually verified for correctness by the first author.

\section{Ethics Statement}

The listening study in this paper was approved by Queen Mary University of London's ethics review board (ref. QMERC20.565.DSEECS25.027). Participation was voluntary and based on informed consent: participants were informed of the purpose, duration, and nature of the task before taking part, and received no compensation. No personally identifiable information was collected, and all responses were stored and analyzed anonymously. 

Our models are trained on the LakhMIDI dataset, which is publicly available and widely used in symbolic music research. It may include transcriptions or arrangements of commercially released, copyrighted compositions. We use this data strictly for non-commercial academic research, while acknowledging that models trained on such data may reproduce fragments of their training examples. We urge practitioners deploying such models in real-world settings to ensure appropriate licensing and to respect the rights of original creators.

\bibliography{ISMIRtemplate}

%
%
%
%

\end{document}